\documentclass[submission,copyright,creativecommons]{eptcs}
\providecommand{\event}{Linearity \& TLLA 2018}

\usepackage{breakurl}
\usepackage{underscore} 
\usepackage{symbols}
\usepackage{todonotes}

\title{The ILLTP Library for Intuitionistic Linear Logic}
\author{
    Carlos Olarte
    \institute{Universidade Federal do Rio Grande do Norte, Brazil}
    \email{carlos.olarte@gmail.com}
    \and
    Valeria de Paiva
    \institute{Nuance Communications, USA}
    \email{valeria.depaiva@gmail.com}
    \and
    Elaine Pimentel\thanks{Olarte and Pimentel are supported by CAPES, CNPq and the project FWF START Y544-N23.}
    \institute{Universidade Federal do Rio Grande do Norte, Brazil}
    \email{elaine.pimentel@gmail.com}
    \and
    Giselle Reis\thanks{This paper was made possible by grant NPRP 7-988-1-178
    from the Qatar National Research Fund (a member of the Qatar Foundation).
    The statements made herein are solely the responsibility of the authors.}
    \institute{Carnegie Mellon University, Qatar}
    \email{giselle@cmu.edu}
}

\def\titlerunning{Linear Logic}
\def\authorrunning{C. Olarte, V. de Paiva, E. Pimentel \& G. Reis}

\begin{document}

\maketitle

\begin{abstract} 
Benchmarking automated theorem proving (ATP) systems using standardized problem
sets is a well-established method for measuring their performance. However, the
availability of such libraries for non-classical logics is very limited.  In
this work we propose a library for benchmarking Girard's  (propositional)
intuitionistic linear logic.  For a quick bootstrapping of the collection of
problems, and for discussing the selection of relevant problems and
understanding their meaning as linear logic theorems, we use translations of the
collection of Kleene's intuitionistic theorems in the traditional monograph
``Introduction to Metamathematics''.  We analyze four different translations of
intuitionistic logic into linear logic and compare their proofs using a linear
logic based prover with focusing. In order to enhance the set of problems in our
library, we apply the three provability-preserving translations to the
propositional benchmarks in the ILTP Library.  Finally, we  generate  a
comprehensive set of reachability problems for Petri nets and encode such
problems as linear logic sequents, thus enlarging our collection of problems.
\end{abstract}

\section{Introduction}\label{sec:intro}

Benchmarking automated theorem proving (ATP) systems using standardized problem
sets is a well-established method for measuring their performance. However, the
availability of such libraries for \textit{non-classical} logics is very
limited. For intuitionistic logic,  several small collections of formulas have
been published and used for testing ATP systems. Raths, Otten and
Kreitz~\cite{iltp} consolidated and extended these small sets to provide the
ILTP Library\footnote{\url{http://iltp.de/}} for first-order and propositional
intuitionistic logic.  For modal systems there are at least  two libraries of theorems~\cite{qmltp1,qmltp2}.

In this paper, we provide a library for benchmarking  Girard's 
linear logic~\cite{girard87tcs}. 
Linear logic is a substructural logic that is a refinement of classical and
intuitionistic logic, combining the dualities of the former with many of the
constructive properties of the latter. Ideas from linear logic have been
influential in fields such as programming languages, game semantics, quantum
physics, as well as linguistics, particularly because of its emphasis on
resource-boundedness, duality, and interaction. In particular, 
linear logic has had an important role as a logical framework for specifying and
reasoning about logical and computational systems (the list is long; some
examples are~\cite{cervesato02ic,MillerP13,ChaudhuriR15,OlarteP17}).
As a consequence, several provers have been built for linear logic for
different purposes (listing some:
Sympli \footnote{\url{https://github.com/chaudhuri/sympli}}, 
llprover\footnote{\url{http://bach.istc.kobe-u.ac.jp/llprover/}},
LL prover\footnote{\url{https://github.com/wujuihsuan2016/LL_prover}}, 
linTAP\footnote{\url{http://www.leancop.de/lintap/}},
LL prover explorer\footnote{\url{https://github.com/andykitchen/linear-logic}},
Lolli\footnote{\url{http://www.lix.polytechnique.fr/~dale/lolli/}}, 
Alcove\footnote{\url{http://cic.puj.edu.co/~caolarte/alcove2/}}).
However, so far, there has been no discussion about the {\em efficiency} or {\em
adequacy} of these provers. In this work we present a comprehensive collection of
propositional tests. 
Moreover, with the aim of comparing proofs obtained from different translations of intuitionistic problems, we use a prototypical prover, 
based on focusing,  to run some experiments.

When designing a benchmark, one has to carefully decide on a set of formulas
that is {\em meaningful} in, at least, three ways:  (1) the formulas should be
able to distinguish several different characteristics of the logical systems and
provers; (2) the set should contain important theorems and paradigmatic formulas
(non necessarily provable); and (3) the set should be large enough, so to serve
as a comparison for different provers and systems.
We first concentrate on (1) and (2), comparing translations of a (small) set of
intuitionistic formulas into linear logic. This will enable the discussion of other possible approaches to follow. We will then deal with (3), by introducing a
library with more than 4.000 problems, including different translations of
intuitionistic formulas as well as translations of reachability problems in
Petri nets as linear logic sequents. 

Starting with (1), it turns out that propositional linear logic ($\LL$) has many aspects that need to be
considered. For example, one could adopt its classical ($\CLL$) or
intuitionistic ($\ILL$) versions. Hence one important task would be to determine
the difference in provability between them, and this is already far from 
trivial. While it is possible to  differentiate the syntax of formulas and the
presentation of the inference rules by the standard restriction on the right
context to having at most one formula in $\ILL$\footnote{We note that in the
literature there are two versions of $\ILL$, having {\em at most} or {\em
exactly} one formula in the right context. This is similar to the problem of
considering intuitionistic/minimal logics. Since in this work we will use a
multiplicative fragment of $\ILL$, we adopted  the version of $\ILL$ having
$\bot$ in the grammar.},  $\FILL$~\cite{FILL} is a
multiple-conclusion system with the same connectives and rules as  $\CLL$, but
restricting the form of the application of such rules. Restricting ourselves to
formulas with the same syntax in classical and intuitionistic versions of $\LL$,
the first interesting question would be which formulas are provable in $\CLL$ but
not in $\ILL$.  This is the same issue \eg\ when building a benchmark for
intuitionistic logic versus the existing ones for classical logic.  But the
linear case is far more complicated, since the lack of the structural rules of
weakening and contraction in both $\ILL$ and $\CLL$ makes these systems
``closer'' to each other than in the case of classical and intuitionistic
logics. Indeed, only very recently~\cite{Laurent18}  the first conservativity
results presented in~\cite{Schellinx91} were generalized. 

Another important aspect to be taken into consideration is {\em
focusing}~\cite{andreoli92jlc}. It turns out that both $\ILL$ and $\CLL$ admit complete 
focused proof systems, and provers can be built using proof search strategies
based on this discipline, which reduces the proof search space. This has an
immediate effect on the proposal of formulas composing a possible benchmark library,
since the amount of exponentials in a formula can make a significant difference
on the performance of provers.

Concerning (2), there is no consensus in the community on a set of ``principal''
theorems that should be used as a test for $\LL$-based theorem provers.  In this
work, we will start the discussion by considering
the translation of a fragment of KleeneÕs basic list  of  intuitionistic logic
 ($\IL$). The first challenge is to
understand {\em how} these intuitionistic theorems should be interpreted in
$\LL$. A first answer would be: use one of the well known translations of $\IL$
into $\LL$. This naive approach has, at least, two problems. First,  it is not
adequate to elect {\em one} translation, since different translations have very
different computational behaviors, as it will be clear in
Section~\ref{sec:trans}. Second, some translations would not give the best
interpretation of linear logic formulas. As a simple example, $A\iimp A$ should
most probably be translated as $A\limp A$, without exponentials since this is
equivalent, as a theorem, to the identity. But {\em any} sound translation from
$\IL$ to $\LL$ adds exponentials to implicational formulas. Hence none of them would
preserve the axiom's interpretation.

To achieve desiderata (3),
we start by applying four different translations from $\IL$ to $\LL$ using
Kleene's collection of $\IL$ theorems.  One of the translations is not validity
perserving, resulting in 22 non-provable formulas.  We propose provable versions
(not following any systematic  translation) for those.  This gives a initial set
of 
271 sequents for our library.
Next, we apply to all the propositional formulas of the ILTP library the three
provability preserving translations considered in this work.
This whole set will not only provide some interesting insights on different
behaviors of $\LL$ formulas coming from different translations in the
literature, but also present a significant set of 1.086
formulas for
benchmarking linear logic based provers.
Finally, to enlarge the number of problems in our library, we have generated automatically several reachability problems from the large collection of Petri nets\footnote{We thank Frank Pfenning, who suggested the use of Petri nets during the TLLA/Linearity 2018 meeting.}, available in the Petri Nets Repository\footnote{\url{https://pnrepository.lip6.fr/}}. 
Then, by using  the standard translation of Petri nets as $\ILL$
formulas~\cite{DBLP:conf/ctcs/Marti-OlietM89,DBLP:conf/agp/Cervesato95}, we
add to our benchmark 3.137  formulas of different levels of difficulty. All
these 
4.494
formulas will constitute our ILLTP library for intuitionistic
linear logic, available at:

\begin{center}
\url{https://github.com/meta-logic/lltp}.
\end{center}

{\em Outline.}  The rest of the paper is organized as follows.
Section~\ref{sec:ll} presents $\LL$, focusing, translations and decorations;
Section~\ref{sec:kleene} explains the different parts that make ILLTP, and some
experiment results;
Section~\ref{sec:conc} concludes the paper and presents some future research
directions.

\section{Linear Logic}\label{sec:ll}

Although we assume that the reader is familiar with linear logic, we review some
of its basic proof theory (see~\cite{troelstra92csli} for more details). 

Linear logic is a substructural logic proposed by Girard~\cite{girard87tcs} as a
refinement of classical and intuitionistic logic.
Formulas for
propositional linear logic ($\LL$) are built from the following grammar
\[
 F ::= p  \mid \one \mid \zero \mid\top \mid \bot  \mid F\tensor F \mid F \lpar 
  F \mid  F\with F
  \mid F \oplus F \mid 
  F\limp F
  \mid \quest F \mid \bang F
\]
where atomic formulas $p$ or their negations $p^\perp=p\limp\bot$ are called {\em
literals}.  The logical connectives for $\LL$  can be divided into the following
groups: the {\em multiplicative} version of conjunction, true, disjunction, and
false, which are written as $\otimes$, $\one$, $\lpar$, $\bottom$, respectively;
and the {\em additive} version of these connectives, which are written as
$\with$, $\top$, $\oplus$, $\zero$, respectively; and the {\em exponentials}
$\bang$ and $\quest$.  

In this work we will concentrate on intuitionistic linear logic
($\ILL$)~\cite{girard87tcs}, having sequents of the form $\Seq{\Gamma}{\Delta}$
where $\Delta$ has at most one formula and $\Gamma$ is a multiset of formulas
built from the grammar above, except for the connective $\lpar$ and the exponential
$\quest$.  The rules of $\ILL$ are depicted in Figure~\ref{fig:ill}.

\begin{figure}[h]
 \hrule
  \vspace{0.1cm}
$$\frac{}{\Seq{p}{p}}\ \init
\qquad
\frac{}{\Seq{}{\one}}\ \one_R
  \qquad
   \frac{}{\Seq{\Gamma}{\top}}\ \top_R
\qquad
\frac{\Seq{\Gamma}{}}{\Seq{\Gamma}{\bottom}}\ \bot_R
\qquad
\frac{\Seq{\Gamma}{\Delta}}{\Seq{\Gamma,\one}{\Delta}}\ \one_L
\qquad
\frac{}{\Seq{\bottom\,}{}}\ \bottom_L
 \qquad
   \frac{}{\Seq{\Gamma,\zero}{\Delta}}\ \zero_L
$$
$$
 \frac{\Seq{\Gamma}{F}\quad \Seq{\Gamma}{G}}
       {\Seq{\Gamma}{F\with G}}
     \ \with_R
      \qquad
   \frac{\Seq{\Gamma_1}{F}\quad \Seq{\Gamma_2}{G}}
       {\Seq{\Gamma_1,\Gamma_2}{F\otimes G}} \ \otimes_R
       \qquad
\frac{\Seq{\Gamma}{F_i}}{\Seq{\Gamma}{F_1\oplus F_2}}
  \ \oplus_{R_i}
    \qquad
  \frac{\Seq{\Gamma,F_i}{\Delta}}{\Seq{\Gamma,F_1\with F_2}{\Delta}}
  \ \with_{L_i}
$$
$$
    \frac{\Seq{\Gamma,F,G}{\Delta}}{\Seq{\Gamma,F\otimes G}{\Delta}}\ \otimes_L
   \qquad
  \frac{\Seq{\Gamma,F}{\Delta}\quad \Seq{\Gamma,G}{\Delta}}
       {\Seq{\Gamma,F\oplus G}{\Delta}}
     \ \oplus_L  
      \qquad
\frac{\Seq{\Gamma,F}{G }}{\Seq{\Gamma}{F\limp G}}
              \ \limp_R
              \qquad
  \frac{\Seq{\Gamma_1}{F}\quad \Seq{\Gamma_2,G}{\Delta}}
       {\Seq{\Gamma_1,\Gamma_2,F\limp G}{\Delta}} \ \limp_L
$$
$$
  \frac{\Seq{\bang\Gamma}{F}}{\Seq{\bang\Gamma}{\bang F}}\ !_{R}
\qquad
  \frac{\Seq{\Gamma,\bang F,\bang F}{\Delta}}{\Seq{\Gamma,\bang F}{\Delta}}
              \ \mathsf{cont}_L
              \qquad
    \frac{\Seq{\Gamma}{\Delta}}{\Seq{\Gamma,\bang F}{\Delta}}
              \ \mathsf{weak}_L
                \qquad
    \frac{\Seq{\Gamma,F}{\Delta}}{\Seq{\Gamma,\bang F}{\Delta}}
              \ \mathsf{der}_L
$$
 \hrule
\caption{System $\ILL$. In $\oplus_{R_i}$ and $\with_{L_i}$, $F_i\ \in\{F_1,F_2\}$. }\label{fig:ill}
\end{figure}

\subsection{Focusing}\label{sec:focusing}

\newcommand{\rfoc}[1]{-_{#1}\!\!\to}
\newcommand{\lfoc}[1]{\xrightarrow{#1}}

Andreoli introduced in~\cite{andreoli92jlc}  a notion of normal form for
cut-free derivations in linear logic.  The connectives of $\LL$ can be divided
into two classes: {\em negative} ($\lpar$, $\bottom$, $\with$, $\top$, and
$\quest$) and {\em positive} ($\otimes$, $\one$, $\oplus$, $\zero$, and
$\bang$). Note that the dual of a negative connective is positive and
vice-versa.  In general, the right introduction rules for negative connectives
are invertible, meaning that the conclusion of any of these introduction rules
is equivalent to its premises. The right introduction rules for the positive
connectives are not necessarily invertible, and may require a choice or a
context restriction on the application of rules.  The notions of negative and
positive polarities are extended to formulas in the natural way by considering
the outermost connective, \eg, $A \oplus B$ is positive while $A \with B$ is
negative. Any bias can be assigned to atomic  $\LL$ formulas. 

A focused proof is organized around two ``phases'' of proof construction: the
{\em negative phase} for introducing negative connectives on the succedent or
positive ones in the antecedent, and the {\em positive phase} for the positive
connectives on the succedent or negative ones in the antecedent.  In the
focusing discipline, formulas are decomposed eagerly in the negative phase,
until only positive formulas on the succedent or negative on the antecedent are
left. Then one of them is non-deterministically chosen to be focused on.

The focused system $\ILLF$ for intuitionistic linear logic is presented in
Figure~\ref{fig:illf}.  There are three kinds of arrows in this proof system and
a pair of contexts to the left of the arrows: 
\begin{itemize}
  \item $\Theta$ is a set of antecedent formulas whose main connective is a
  bang, being hence the unbounded (classical) context;
  \item $\Gamma$ is a multiset of linear formulas, behaving as the bounded (linear)
  context;
  \item sequents of the form $\Theta:\Gamma\rfoc{F}$ or
  $\Theta:\Gamma\lfoc{F}\Delta$ belong to the positive phase and introduce the
  logical connective of the ``focused'' formula $F$: building proofs of such
  sequents may require non-invertible proof steps to be taken;
  \item sequents of the form $\Theta:\Gamma \longrightarrow \Delta$ belong to
  the negative phase and decompose the multisets $\Gamma,\Delta$ in such a way
  that only inference rules over negative formulas on the right or positive ones
  on the left are applied.
\end{itemize}
The structural rules $D_{L1} , D_{L2}$ and $D_R$ make the transition between
negative and positive phases.  The \emph{positive} phase begins by choosing  a non-atomic fomula $N_a$ from the classical context or a
negative left formula $N$ (resp.  a positive right formula $P$) on which to
focus using $D_{L1} , D_{L2}$ (resp.  $D_R$). Atoms in $\ILLF$ always have a
{\em positive} bias\footnote{Although $\ILL$ can allow any polarity assignment for atomic formulas, we will present here the system that was actually implemented.}. Positive rules are applied to $N, N_a/P$ until either
$\one$ or an  atom is encountered in the right, or $\bot$ is on the left (and,
in such cases, the proof must end by applying the respective axiom), or the
promotion rule ($\bang$) is applied,  or yet a  positive left  (respect.
negative right) subformula is encountered,  when the proof switches to the
negative phase by applying $R_L$ (resp. $R_R$).

An observation is in order: the system presented in Figure~\ref{fig:illf} induces a {\em weak focusing}, in the sense that, a priori, the positive phase could start even with the presence, in one of the contexts, 
of formulas amenable to negative rule applications. However, our rewrite-based implementation (see Section~\ref{sec:l-kleene}) forces the strong focusing approach:  rules in the negative phase are exhaustively applied until a normal form is found, i.e., in all the resulting subgoals, 
there are no positive formulas in the linear context, nor a negative formula in the succedent. 

\begin{figure}[h]
\hrule
\vspace{.2cm}
\begin{center}
 \textsc{\bf Negative Phase}
\end{center}
\[
\infer[\tensor_L]{\Theta:\Gamma, F \tensor G \longrightarrow \Delta}
{
\Theta:\Gamma, F,  G \longrightarrow \Delta
}
\quad
\infer[\limp_R]{\Theta:\Gamma \longrightarrow F \limp G}
{
\Theta:\Gamma, F \longrightarrow G
}
\quad
\infer[\one_L]{\Theta:\Gamma, \one \longrightarrow \Delta}
{\Theta:\Gamma \longrightarrow \Delta}
\quad
\infer[\bot_R]{\Theta:\Gamma \longrightarrow \bot}
{\Theta:\Gamma \longrightarrow }
\]
\[\infer[\top_R]{\Theta:\Gamma \longrightarrow \top}{}
\quad
\infer[0_L]{\Theta:\Gamma,0 \longrightarrow \Delta}{}
\quad
\infer[\bang_L]{\Theta:\Gamma, \bang F \longrightarrow \Delta}
{\Theta,F:\Gamma \longrightarrow \Delta}
\]
\[
\infer[\with_R]{\Theta:\Gamma\longrightarrow F
\with G}
{
\Theta:\Gamma\longrightarrow F
&
\Theta:\Gamma\longrightarrow G}
\quad
\infer[\oplus_L]{\Theta:\Gamma,F\oplus H\longrightarrow \Delta}
{
\Theta:\Gamma,F\longrightarrow  \Delta
&
\Theta:\Gamma,H\longrightarrow  \Delta
}
\]
\begin{center}
 \textsc{\bf Positive Phase}
\end{center}
\[
\infer[\tensor_R]{\Theta:\Gamma_1,\Gamma_2\rfoc{F
\tensor G}}
{
\Theta:\Gamma_1 \rfoc{F}
&
\Theta:\Gamma_2 \rfoc{G}
}
\qquad 
\infer[\limp_L]{\Theta:\Gamma_1,
\Gamma_2\lfoc{F \limp G} \Delta}
{
 \Theta:\Gamma_1\rfoc{F} 
&
\Theta:\Gamma_2 \lfoc{G} \Delta
}
\quad
\infer[\oplus_{R_i}]{\Theta: \Gamma\rfoc{F_1
\oplus F_2}}
{
\Theta:  \Gamma\rfoc{F_i}
}
\]
\[
\infer[\with_{L_i}]{\Theta: \Gamma \lfoc{F_1
\with F_2}C}
{
\Theta:  \Gamma\lfoc{F_i}\Delta
}
\quad
\infer[1_R]{\Theta: \cdot \;\rfoc{1}
}{}
\quad 
\infer[\bot_L]{\Theta: \cdot \;\lfoc{\bot}
}{}
\quad 
\infer[\bang_R]{\Theta: \cdot \; \rfoc{\bang
F}}
{\Theta: \cdot \;\longrightarrow F}
\]
\vspace{0.2pt}
\[
\infer[I_R \hbox{ given $p \in (\Theta \cup
\Gamma)$ and $\Gamma  \subseteq \{p\} $}]{\Theta: \Gamma\rfoc{p}
}{}
\]

\begin{center}
 \textsc{\bf Structural Rules}
\end{center}
\vspace{-2mm}
\[
\infer[D_{L1}]{\Theta,N_a:\Gamma \longrightarrow
\Delta}
{\Theta,N_a:\Gamma \lfoc{N_a} \Delta}
\quad
\infer[D_{L2}]{\Theta:\Gamma,N \longrightarrow
\Delta}
{\Theta:\Gamma \lfoc{N} \Delta}
\quad 
\infer[D_R]{\Theta:\Gamma \longrightarrow
P}
{\Theta:\Gamma \rfoc{P}}
\]
\[
\infer[R_L]{\Theta:\Gamma \lfoc{P} \Delta}
{
\Theta: \Gamma,P \longrightarrow \Delta
}
\quad
\infer[R_R]{\Theta: \Gamma\rfoc{N} }
{
\Theta: \Gamma\longrightarrow N}
\]
\hrule
\caption{System $\ILLF$: a focused proof system for $\ILL$. Here, $p$ is an
atomic formula (we will assume positive bias for atoms); $N$ is a negative formula; $P$ is a positive formula; $N_a$ is a
non-atomic formula. In $\oplus_{R_i}$ and $\with_{L_i}$, $F_i\ \in\{F_1,F_2\}$.
}
\label{fig:illf}
\end{figure}

Finally, we observe that, although focusing is a complete strategy that
decreases considerably the search space, it is not enough for avoiding loops on
proof search. Let us give two simple examples. A focused proof of the sequent
$\cdot: \bang A,\bang(\bang A\limp B)\longrightarrow \zero$ starts by storing
$A, \bang A\limp B$ in the classical context and then it focuses on $\bang A\limp
B$ continuously, adding various copies of $B$ in the linear context. Since the
sequents produced in this process are all different, no loop detection will work
for this case. One way of avoiding this kind of behavior would be by restricting
the number of focusing steps over the same formula in the classical context.
This is a sound, but not complete, strategy. Another way to deal with this
problem  would be to  introduce a heuristics  in order to stop the computation
or to modify the sequents, depending on the shape of formulas in the context.
In this particular example, it is enough to observe that sequents of the form
$\Theta,A,A\limp B:\Gamma \longrightarrow C$ can be re-written as
$\Theta,A,B,A\limp B:\Gamma \longrightarrow C$, and continue by analyzing
the formula $B$. 

Another pathological behavior may arise with a sequent of the shape $\Theta, P_1
\with P_2 : \Gamma \longrightarrow C$ where $P_i$ is a positive formula. In
this case, if we decide to focus on the formula $P_1 \with P_2$, we are forced
to choose one of the $P_i$ and then, the positive phase ends by adding yet
another copy of $P_i$ into the linear context.  In this particular case, we may
use the well known equivalence $!(F \with G) \equiv ! F \otimes !  G$ and
rewrite the above sequent into a simpler one: $\Theta, P_1 , P_2 : \Gamma
\longrightarrow C$. If $P_1$ and $P_2$ are atomic formulas, this
simplification is quite useful since  those atoms will remain in the classical
context (remember that it is not possible to focus on them) and they will never be added
to the linear context.

In a broader sense, the analysis of proof theoretic properties of $\LL$  must guide the design of more advanced/efficient proof search procedures. 
For instance, invertibility and permutability of rules 
lead to systems that reduce the non-determinism \cite{DBLP:conf/lfcs/GalmicheP94,andreoli92jlc} as already mentioned. Moreover,  the resource management problem has been tackled  
by delaying, as much as possible, the decision on how to split the linear context \cite{DBLP:journals/tcs/CervesatoHP00,DBLP:conf/lpar/LopezP99}. On the other hand, systems for forward reasoning 
 have been also explored in the context of $\LL$ (see e.g., \cite{DBLP:journals/jolli/Mints93,DBLP:journals/jar/Tammet94,DBLP:conf/kgc/Tammet97,DBLP:conf/cade/ChaudhuriP05}),
 usually showing a better performance that those based on 
backward reasoning \cite{DBLP:journals/jar/Tammet94,DBLP:conf/cade/ChaudhuriP05}. Since our goal with this work is to present an initial library for $\LL$
and not to test particular theorem provers or automated reasoning techniques, we will not focus our attention on
these issues. We just note that no silver-bullet will work in all the
cases since, as it is well known, propositional linear logic is
undecidable~\cite{DBLP:journals/apal/LincolnMSS92}.

\subsection{Translations and Decorations}\label{sec:trans} 

A naive approach for building a set of test formulas for $\LL$ based provers
would be to use one of the well known translations of intuitionistic (or
classical) formulas into $\LL$. Since there are several ways of translating a
formula from $\IL$ to $\LL$, we asked ourselves which one would be the
best option, if any.
Each translation characterizes a different linear view of intuitionistic
formulas and it is interesting and relevant to establish a comparison between
them.
We analyze four: a multiplicative translation, 
the original  Girard's translation, Girard's positive translation and Miller and
Liang's $0/1$ translation. 

The multiplicative translation trivially substitutes the intuitionistic
connectives by their multiplicative linear version
$$
\begin{array}{rclcrclcrcl}
(p)^m & = & p & \qquad & (A\iimp B)^m & = & A^m\limp B^m&\qquad&
(A\wedge B)^m & = & A^m\tensor B^m\\
(t)^m & = & \one & & (A\vee B)^m & = & A^m \lpar B^m & & 
(f)^m & = & \bottom
\end{array}
$$
Translation of sequents is given by $(\seq{\Gamma}{A})^m=\seq{\Gamma^m}{A^m}$.
Observe that this translation {\em does not} preserve provability: for instance,
diagonals $A\otimes A\to A$ exist  in $\IL$, but not in $\LL$.

Girard's translation~\cite{girard87tcs}, also known as {\em call-by-name} translation,
is the most well known translation of $\IL$ into $\LL$, defined as 
$$
\begin{array}{rclcrclcrcl}
(p)^g & = & p & \qquad & (A\iimp B)^g & = & \bang A^g\limp B^g&\qquad&
(A\wedge B)^g & = & A^g\with B^g\\
(t)^g & = & \top & & (A\vee B)^g & = & \bang A^g\oplus \bang B^g & & 
(f)^g & = & \zero
\end{array}
$$
Sequents are translated as $(\seq{\Gamma}{A})^g=\seq{\bang \Gamma^g}{A^g}$.
Girard's translation preserves provability but is not a {\em decoration} in the
sense of~\cite{DanosJS95}, namely, a proof of $A$ in $\IL$ is transformed into
a proof of $A^g$ in $\LL$ which is not isomorphic to the original one.

Girard proposed in the same paper~\cite{girard87tcs} another translation,  known
as {\em call-by-value} translation. Henceforth, we will call this translation
\emph{positive}, since $\IL$ formulas become positive $\LL$ formulas.
$$
\begin{array}{rclcrclcrcl}
(p)^p & = & \bang p & \qquad & (A\iimp B)^p & = & \bang (A^p\limp B^p)&\qquad&
(A\wedge B)^p & = & A^p\tensor B^p\\
(t)^p & = & \one & & (A\vee B)^p & = & A^p\oplus  B^p & & 
(f)^p & = & \zero
\end{array}
$$
Sequents are translated as $(\seq{\Gamma}{A})^p=\seq{\Gamma^p}{A^p}$. 
It is easy to see that the positive translation is a decoration: proofs of $A^p$ in $\LL$ are
isomorphic to proofs of $A$ in $\IL$ (see~\cite{DanosJS95} for details).

Another interesting translation is the $0/1$ translation~\cite{LiangM09}, which
distinguishes the polarity of formulas in a sequent.
$$
\begin{array}{rclcrclcrcl}
(p)^0 & = & p & \qquad & (A\iimp B)^0 & = & \bang A^1\limp \bang B^0&\qquad&
(A\wedge B)^0 & = & \bang A^0\with\bang B^0\\
(t)^0 & = & \top & & (A\vee B)^0 & = & \bang A^0\oplus  \bang B^0 & & 
(f)^0 & = & \zero\\ \\
(p)^1 & = & p & \qquad & (A\iimp B)^1 & = & \bang (\bang A^0\limp B^1)&\qquad&
(A\wedge B)^1 & = & \bang (A^1\with B^1)\\
(t)^1 & = & \one & & (A\vee B)^1 & = & \bang A^1\oplus  \bang B^1 & & 
(f)^1 & = & \zero
\end{array}
$$
\vspace{-0.2cm}

The translation of sequents is given by $(\seq{\Gamma}{A})^{0/1}=\seq{\bang\Gamma^0}{A^1}$. 
Using this translation, {\em focused proofs} in $\ILLF$ are in bijective
correspondence with proofs in $\IL$. In a loose sense, this can be considered a
decoration if the isomorphism is interpreted ``modulo focusing''.
In the focusing context, this is referred to as \emph{adequacy on the level of derivations}~\cite{Nigam2010}.

Basically these four translations differ on their use of bangs and
polarization of atoms. The multiplicative translation introduces no
bangs; Girard's translation forces 
backchaining proofs; the positive translation sets the global preference to
be forward-chaining and atoms have positive polarity; the $0/1$ translation
is asymmetric, and it does not impose restrictions on atoms. 
Of course, less bangs implies shorter proofs. For instance, consider the $\IL$
sequent $A \iimp B, B \iimp C \vdash A \iimp C$ and the corresponding proofs using  
the four translations in Figure \ref{fig:trans}.

\begin{figure}
\begin{center}
\begin{tabular}{c}
\hline
Multiplicative translation\\\hline\\
$\infer[\Rneg]{\LLSec{ \LLempty }{A \LLimp B , B \LLimp C }{A \LLimp C}}{\infer[\RLimp]{\LLSec{ \LLempty }{A , A \LLimp B , B \LLimp C }{C}}{\infer{\LLSec{ \LLempty }{A }{A}}{} & \infer[\RLimp]{\LLSec{ \LLempty }{B , B \LLimp C }{C}}{\infer{\LLSec{ \LLempty }{B }{B}}{} & \infer{\LLSec{ \LLempty }{C }{C}}{} } } }$\\\\\hline
Call-by-name\\\hline\\
\resizebox{.9\textwidth}{!}{$\infer[\Rneg]{\LLSec{ \LLempty }{ \LLbang ( \LLbang (A) \LLimp B) , \LLbang ( \LLbang (B) \LLimp C) }{ \LLbang (A) \LLimp C}}{\infer[\RdecideC]{\LLSec{A , \LLbang (A) \LLimp B , \LLbang (B) \LLimp C }{ \LLempty }{C}}{\infer[\RLimp]{\LLSec{A , \LLbang (A) \LLimp B , \LLbang (B) \LLimp C }{ \LLbang (B) \LLimp C }{C}}{\infer[\Rbang]{\LLSec{A , \LLbang (A) \LLimp B , \LLbang (B) \LLimp C }{ \LLempty }{ \LLbang (B)}}{\infer[\RdecideC]{\LLSec{A , \LLbang (A) \LLimp B , \LLbang (B) \LLimp C }{ \LLempty }{B}}{\infer[\RLimp]{\LLSec{A , \LLbang (A) \LLimp B , \LLbang (B) \LLimp C }{ \LLbang (A) \LLimp B }{B}}{\infer[\Rbang]{\LLSec{A , \LLbang (A) \LLimp B , \LLbang (B) \LLimp C }{ \LLempty }{ \LLbang (A)}}{\infer{\LLSec{A , \LLbang (A) \LLimp B , \LLbang (B) \LLimp C }{ \LLempty }{A}}{} } & \infer{\LLSec{A , \LLbang (A) \LLimp B , \LLbang (B) \LLimp C }{B }{B}}{} } } } & \infer{\LLSec{A , \LLbang (A) \LLimp B , \LLbang (B) \LLimp C }{C }{C}}{} } } }$}
\\\\\hline
Call-by-value\\\hline\\
\resizebox{.95\textwidth}{!}{ $\infer[\Rneg]{\LLSec{ \LLempty }{ \LLbang ( \LLbang (A) \LLimp \LLbang (B)) , \LLbang ( \LLbang (B) \LLimp \LLbang (C)) }{ \LLbang ( \LLbang (A) \LLimp \LLbang (C))}}{\infer[\Rbang]{\LLSec{ \LLbang (A) \LLimp \LLbang (B) , \LLbang (B) \LLimp \LLbang (C) }{ \LLempty }{ \LLbang ( \LLbang (A) \LLimp \LLbang (C))}}{\infer[\Rneg]{\LLSec{ \LLbang (A) \LLimp \LLbang (B) , \LLbang (B) \LLimp \LLbang (C) }{ \LLempty }{ \LLbang (A) \LLimp \LLbang (C)}}{\infer[\RdecideC]{\LLSec{A , \LLbang (A) \LLimp \LLbang (B) , \LLbang (B) \LLimp \LLbang (C) }{ \LLempty }{ \LLbang (C)}}{\infer[\RLimp]{\LLSec{A , \LLbang (A) \LLimp \LLbang (B) , \LLbang (B) \LLimp \LLbang (C) }{ \LLbang (B) \LLimp \LLbang (C) }{ \LLbang (C)}}{\infer[\Rbang]{\LLSec{A , \LLbang (A) \LLimp \LLbang (B) , \LLbang (B) \LLimp \LLbang (C) }{ \LLempty }{ \LLbang (B)}}{\infer[\RdecideC]{\LLSec{A , \LLbang (A) \LLimp \LLbang (B) , \LLbang (B) \LLimp \LLbang (C) }{ \LLempty }{B}}{\infer[\RLimp]{\LLSec{A , \LLbang (A) \LLimp \LLbang (B) , \LLbang (B) \LLimp \LLbang (C) }{ \LLbang (A) \LLimp \LLbang (B) }{B}}{\infer[\Rbang]{\LLSec{A , \LLbang (A) \LLimp \LLbang (B) , \LLbang (B) \LLimp \LLbang (C) }{ \LLempty }{ \LLbang (A)}}{\infer{\LLSec{A , \LLbang (A) \LLimp \LLbang (B) , \LLbang (B) \LLimp \LLbang (C) }{ \LLempty }{A}}{} } & \infer[\Rneg]{\LLSec{A , \LLbang (A) \LLimp \LLbang (B) , \LLbang (B) \LLimp \LLbang (C) }{ \LLbang (B) }{B}}{\infer{\LLSec{A , B , \LLbang (A) \LLimp \LLbang (B) , \LLbang (B) \LLimp \LLbang (C) }{ \LLempty }{B}}{} } } } } & \infer[\Rneg]{\LLSec{A , \LLbang (A) \LLimp \LLbang (B) , \LLbang (B) \LLimp \LLbang (C) }{ \LLbang (C) }{ \LLbang (C)}}{\infer[\Rbang]{\LLSec{A , C , \LLbang (A) \LLimp \LLbang (B) , \LLbang (B) \LLimp \LLbang (C) }{ \LLempty }{ \LLbang (C)}}{\infer{\LLSec{A , C , \LLbang (A) \LLimp \LLbang (B) , \LLbang (B) \LLimp \LLbang (C) }{ \LLempty }{C}}{} } } } } } } }$}
\\\\\hline
0/1\\\hline\\
\resizebox{.95\textwidth}{!}{ $\infer[\Rneg]{\LLSec{ \LLempty }{ \LLbang ( \LLbang (A) \LLimp \LLbang (B)) , \LLbang ( \LLbang (B) \LLimp \LLbang (C)) }{ \LLbang ( \LLbang (A) \LLimp C)}}{\infer[\Rbang]{\LLSec{ \LLbang (A) \LLimp \LLbang (B) , \LLbang (B) \LLimp \LLbang (C) }{ \LLempty }{ \LLbang ( \LLbang (A) \LLimp C)}}{\infer[\Rneg]{\LLSec{ \LLbang (A) \LLimp \LLbang (B) , \LLbang (B) \LLimp \LLbang (C) }{ \LLempty }{ \LLbang (A) \LLimp C}}{\infer[\RdecideC]{\LLSec{A , \LLbang (A) \LLimp \LLbang (B) , \LLbang (B) \LLimp \LLbang (C) }{ \LLempty }{C}}{\infer[\RLimp]{\LLSec{A , \LLbang (A) \LLimp \LLbang (B) , \LLbang (B) \LLimp \LLbang (C) }{ \LLbang (B) \LLimp \LLbang (C) }{C}}{\infer[\Rbang]{\LLSec{A , \LLbang (A) \LLimp \LLbang (B) , \LLbang (B) \LLimp \LLbang (C) }{ \LLempty }{ \LLbang (B)}}{\infer[\RdecideC]{\LLSec{A , \LLbang (A) \LLimp \LLbang (B) , \LLbang (B) \LLimp \LLbang (C) }{ \LLempty }{B}}{\infer[\RLimp]{\LLSec{A , \LLbang (A) \LLimp \LLbang (B) , \LLbang (B) \LLimp \LLbang (C) }{ \LLbang (A) \LLimp \LLbang (B) }{B}}{\infer[\Rbang]{\LLSec{A , \LLbang (A) \LLimp \LLbang (B) , \LLbang (B) \LLimp \LLbang (C) }{ \LLempty }{ \LLbang (A)}}{\infer{\LLSec{A , \LLbang (A) \LLimp \LLbang (B) , \LLbang (B) \LLimp \LLbang (C) }{ \LLempty }{A}}{} } & \infer[\Rneg]{\LLSec{A , \LLbang (A) \LLimp \LLbang (B) , \LLbang (B) \LLimp \LLbang (C) }{ \LLbang (B) }{B}}{\infer{\LLSec{A , B , \LLbang (A) \LLimp \LLbang (B) , \LLbang (B) \LLimp \LLbang (C) }{ \LLempty }{B}}{} } } } } & \infer[\Rneg]{\LLSec{A , \LLbang (A) \LLimp \LLbang (B) , \LLbang (B) \LLimp \LLbang (C) }{ \LLbang (C) }{C}}{\infer{\LLSec{A , C , \LLbang (A) \LLimp \LLbang (B) , \LLbang (B) \LLimp \LLbang (C) }{ \LLempty }{C}}{} } } } } } }$ 
}
\end{tabular}
\end{center}
\caption{\label{fig:trans}
Proof of the sequent $A \iimp B, B \iimp C \vdash A \iimp C$ using  the four translations. The formulas in \blue{blue}~    represent
the classical context and the $\star$ rule condenses all the rules in the negative phase. }
\end{figure}

\section{The ILLTP Library}\label{sec:kleene}
The ILLTP library is composed of problems from three sources: Kleene's
``Introduction to Metamathematics'', ILTP (a library of problems for
intuitionistic theorem provers), and 
Petri nets from the Model Checking Contest \footnote{\url{https://pnrepository.lip6.fr/mcc/}}.
We use a syntax similar to TPTP~\cite{Sut17}\footnote{\url{http://www.tptp.org}}, where axioms (formulas on the left side of the
sequent) and conjectures (the formula on the right side of the sequent) are
specified as 
\texttt{fof(name, axiom, F)} and \texttt{fof(name, conjecture, F)} respectively.
The name chosen is not important. The denominations \texttt{axiom} and
\texttt{conjecture} are fixed. \texttt{F} is a linear logic formula built from
the connectives: \texttt{*}, \texttt{\&}, \texttt{+}, \texttt{|}, \texttt{-o},
\texttt{!}, \texttt{?}, corresponding to $\otimes$, $\binampersand$, $\oplus$,
$\bindnasrepma$, $\multimap$, $!$, $?$, respectively\footnote{
Although our library is used here for $\ILL$, the format of our files and translations support $\LL$ formulas as well, so that our library can be extended to the classical case as well.}.

\subsection{Kleene's problems}

KleeneÕs traditional book  ``Introduction to Metamathematics''~\cite{klccne1952im} has a collection of interesting intuitionistic theorems. They are rather straightforward, thus
they would not be especially useful for testing the efficiency of a prover. Instead,
they can be regarded as a minimal set of intuitionistic theorems that a
\emph{sound} prover should be able to derive. As such, they can be valuable to  uncover bugs and  unsoundness. 
Our first goal is to set up a similar set for $\LL$. 

The main challenge is to understand how these intuitionistic theorems should be
interpreted in $\LL$. Deciding whether to translate intuitionistic disjunction
as the  multiplicative disjunction $\lpar$ of linear logic or the additive
disjunction $\oplus$ changes the target system under consideration, thus we
prefer to not consider the intuitionistic disjunction to begin with. Hence we
will start by considering what we call the \textit{rudimentary fragment of $\IL$}, which is
very well-behaved: semantically this fragment corresponds to cartesian closed
categories.
Figure~\ref{fig-kleene} shows the 61 theorems considered. 

Applying each translation defined in Section~\ref{sec:trans} to these 61
sequents gives rise to 244 different $\ILL$ sequents.  As already noted,
provability is not preserved in the multiplicative translation. The reason for
that, other than the obvious absence of structural rules in the left context, is
that the multiplicative false $\bot$ is relevant, so while $0 \vdash B$ for any
$B$,  $A\otimes (A\limp \bot)\not\vdash B$ in $\LL$.  The other three
translations fix this by systematically adding bangs and additive connectives. 

The multiplicative translation of all sequents in Kleene's list is in
Appendix~\ref{app:mult}, including the 22 ones that are not provable (indicated by
$\not\vdash$). In Figure~\ref{fig:alt-enc} we present an alternative translation
for them, using a small number of bangs and/or additives.

\begin{figure}
{\footnotesize
\begin{multicols}{2}
\begin{enumerate}\setlength\itemsep{0em}
\item $\vdash A\to A$ 
\item $A\to B, B\to C\vdash A\to C$ 
\item $A\to (B\to C)\vdash B\to (A\to C)$ 
\item $A\to (B\to C) \vdash A\land B\to C$ 
\item $A\land B\to C\vdash A\to (B\to C)$ 
\item $A\to B\vdash (B\to C) \to (A\to C)$ 
\item $A\to B\vdash (C\to A) \to (C\to B)$ 
\item $A\to B\vdash A\land C\to B\land C$ 
\item $A\to B\vdash C\land  A \to C\land B$ 
\item  $\neg A\vdash A\to B$ 
\item $A\vdash \neg A\to B$ 
\item $B \vdash A\to B$ 
\item $A\to B\vdash \neg B\to \neg A$ 
\item $A\to \neg B\vdash\neg \neg B\to \neg A$ 
\item $A\to B, B\to A\vdash A\sim B$ 
\item $A\sim B\vdash A\to B$
\item $A\sim B\vdash B\to A$
\item $A\sim B, A\vdash  B$
\item $A\sim B, B\vdash  A$
\item $\vdash A\sim A$ 
\item $A\sim B\vdash B\sim A$
\item $A\sim B, B\sim C\vdash A\sim C$
\item $A\to (B\to C), \neg\neg A, \neg\neg B \vdash \neg\neg C$
\item $\neg \neg (A\to B)\vdash \neg \neg A\to \neg\neg B $
\item $\neg \neg (A\to B), \neg \neg (B\to C)\vdash \neg \neg (A\to C)$
\item $\vdash \neg \neg (A\land B) \sim (\neg\neg A\land \neg\neg B)$
\item $\vdash \neg \neg (A\sim B) \sim (\neg\neg (A\to B) \land \neg\neg (B\to A))$
\item $A \sim B \vdash (A \to C) \sim (B \to C)$
\item $A \sim B \vdash (C \to A) \sim (C \to B)$
\item $A \sim B \vdash (A \land C) \sim (B \land C)$
\item $A \sim B \vdash (C \land A) \sim (C \land B)$
\item $A \sim B \vdash \neg A \sim \neg B$
\item $\vdash ((A \land B) \land C) \sim (A \land (B \land C))$
\item $\vdash (A \land B) \sim (B \land A)$ 
\item $\vdash (A \land A) \sim A$ 
\item $A \vdash (A \to B) \sim B$
\item $B \vdash (A \to B) \sim B$
\item $\neg A \vdash (A \to B) \sim \neg A$
\item $\neg B \vdash (A \to B) \sim \neg A$
\item $B \vdash (A \land B) \sim A$
\item $\neg B \vdash (A \land B) \sim B$
\item $\vdash A \to \neg \neg A$
\item$\vdash \neg \neg \neg A \sim \neg A$
\item$\vdash \neg (A \land \neg A)$
\item $\vdash \neg (A \sim \neg A)$
\item $\vdash \neg \neg (\neg \neg A \to A)$
\item $\vdash (A \land (B \land \neg B)) \sim (B \land \neg B)$
\item $\vdash (A \to B) \to \neg (A \land \neg B)$
\item $\vdash (A \to \neg B) \sim (\neg (A \land B))$
\item $\vdash  (\neg (A \land B)) \sim (\neg \neg A \to \neg B)$
\item $\neg \neg B \to B \vdash (\neg \neg A \to B) \sim (A \to B)$
\item $\neg \neg B \to B \vdash (A \to B) \sim (\neg (A \land \neg B))$
\item $\vdash (\neg \neg A \to B) \to \neg (A \land \neg B)$
\item $\vdash (A \land B) \to \neg (A \to \neg B)$
\item $\vdash (A \land \neg B) \to \neg (A \to B)$
\item $\vdash \neg \neg A \land B \to \neg (A \to \neg B)$
\item $\vdash (\neg \neg A \land \neg B) \sim \neg (A \to B)$
\item $\vdash \neg (A \to B) \sim \neg \neg (A \land \neg B)$
\item $\vdash \neg \neg (A \to B) \sim \neg (A \land \neg B) $
\item $\vdash  \neg (A \land \neg B) \sim (A \to \neg \neg B) $
\item $\vdash  (A \to \neg \neg B) \sim (\neg \neg A \to \neg \neg B)$
\end{enumerate}
\end{multicols}
}
\vspace{-0.5cm}
\caption{Kleene's theorems 
collected from~\cite{klccne1952im} (page 113
onwards). Only the  $(\to,\land)$ fragment is considered. Bi-implication is defined as $A\sim B = (A\iimp B)\wedge (B\iimp A)$.
\label{fig-kleene}
}
\end{figure}

\begin{figure}
\begin{multicols}{2}
{\footnotesize
\begin{enumerate}\setlength\itemsep{0em}
\setcounter{enumi}{9}
\item $ A\limp\zero \vdash A\limp B$
\item $ A\vdash (A\limp\zero) \limp B$
\item $B \vdash \bang A\limp B$ 
 \setcounter{enumi}{15}
\item  $(A\limp B)\otimes \bang (B\limp A) \vdash A\limp B$ 
\item $\bang (A\limp B)\otimes (B\limp A) \vdash B\limp A$ 
\item  $A\multimapboth B, A \vdash  B\otimes (B\limp A)$
\item $A\multimapboth B, B\vdash A\otimes (A\limp B)$
\setcounter{enumi}{25}
\item \begin{itemize}
\item[a.] $\vdash  ((A\with B)^{{\bot}{\bot}}) \limp  (A^{{\bot}{\bot}}\with  B^{{\bot}{\bot}})$ 
\item[b.]  $\vdash (A^{{\bot}{\bot}}\otimes  B^{{\bot}{\bot}}) \limp(A\otimes B)^{{\bot}{\bot}}$
\end{itemize}
\item    \begin{itemize}
\item[a.] $\vdash  (\bang(A \limp B)\tensor\bang(B\limp A))^{{\bot}{\bot}} \limp  [(A\limp B)^{{\bot}{\bot}} \with  (B\limp A)^{{\bot}{\bot}}]$
\item[b.] $\vdash (A\limp B)^{{\bot}{\bot}} \otimes  (B\limp A)^{{\bot}{\bot}}\limp (A\multimapboth B)^{{\bot}{\bot}}$
\end{itemize}
\setcounter{enumi}{34}
\item $\vdash (\bang A \otimes \bang A) \multimapboth \bang A$ 
\item $A \vdash ((A \limp B)\limp  B)\tensor (B\limp (\bang A\limp B))$ 
\item $B \vdash (\bang(A \limp B)\limp  B)\tensor (B\limp (\bang A\limp B))$ 
\item  $A^\perp \vdash (\bang (A \limp B)  \limp A^\perp)\tensor ((A\limp 0)\limp  (A \limp  B))  $

\item $ B\limp\zero \vdash (A \limp B)  \multimapboth (A\limp\zero)$

\item $B \vdash ((A \tensor \bang B) \limp A)\otimes ( A\limp(A\tensor B))$ 
\item $B\limp\zero\vdash ((\bang A\tensor B)\limp B)\tensor (B\limp (A\tensor B))$
\setcounter{enumi}{44}
\item $\vdash (\bang(A\limp A^\perp)\otimes((\bang A)^\perp\limp \bang A))^\perp$ 
\item $\vdash (\bang(\bang(A^\perp\limp\zero)) \limp A)^{\perp})^{\perp}$ 
\item $\vdash A \otimes (B \otimes (B\limp\zero)) \multimapboth (B \otimes (B\limp\zero))$
\setcounter{enumi}{56}
\item  \begin{itemize}
\item[a.] $\vdash (A^{\perp\perp} \otimes B^\perp) \limp  (A \limp B)^\perp$
\item[b.] $\vdash (\bang A \limp B)^\perp\limp ((A\limp\zero)^\perp \with B^\perp)$ 
\end{itemize}
\item \begin{itemize}
\item[a.] $\vdash \bang((\bang A \limp B)^\perp)\limp  ((A \otimes B^\perp)\limp\zero)^{\perp}$
\item[b.] $\vdash (A\otimes B^\perp)^{\perp\perp}\limp (A\limp B)^\perp$  
\end{itemize}
\item \begin{itemize}
\item[a.] $\vdash (A \limp B)^{\perp\perp} \limp  (A \otimes B^\perp)^\perp $
\item[b.] $\vdash ((A\tensor B^\perp)\limp\zero) \limp (\bang(\bang A\limp B)^{\perp})^{\perp} $ 
\end{itemize}
\end{enumerate}
}
\end{multicols}
\vspace{-0.5cm}
\caption{Alternative translation for Kleene theorems \label{fig:alt-enc}}
\end{figure}
 
\subsection{ILTP problems}

The problems from the ILTP library for intuitionistic theorem
provers~\cite{iltp} were parsed and translated into linear logic using the three
provability preserving translations presented in Section~\ref{sec:trans}. Due to
the number and size of problems, it is unpractical to apply the multiplicative
translation and find the unprovable sequents to propose an alternative
translation as it was done with the previous set of problems.

The original library contains 274
 problems (including theorems and
non-theorems), separated into three categories: LCL (logical calculi), SYJ and
SYN (syntactic problems that have no obvious interpretation in intuitionistic
and classical logic, respectively).
Their translation resulted in 
822 
linear logic problems.  Moreover, we also considered the remaining Kleene's theorems in ~\cite{klccne1952im} (i.e., those out of the rudimentary fragment) resulting in a  total of  1.086 problems in this collection. 
 
\subsection{Comparing translations}\label{sec:l-kleene}

We specified $\IL$ and different fragments of linear logic in rewriting logic
($\RW$, see e.g.~\cite{DBLP:journals/jlp/Meseguer12}) and implemented in
Maude\footnote{\url{http://maude.cs.uiuc.edu}} a very basic prover for $\IL$  as
well as for $\ILLF$ and $\LLF$. The use of $\RW$ leads to a clear separation
between deterministic inference rules that can be eagerly applied (as those in
the negative phase) and non-deterministic inference rules where backtracking may
be needed (as those in the positive phase).  Moreover, the minimal distance
between the represented logic  ($\IL$, $\ILLF$ and $\LLF$) and its specification
in $\RW$ allowed us to quickly implement a good prototypical tool useful for our
tests. Although more efficient provers can be built by e.g., including
sophisticated heuristics and specialized data structures, our prototypical
implementations were enough to compare the different translations.  

Using these provers, we generated \LaTeX{} proofs for the intuitionistic problems
from Kleene and ILTP, plus for all $\ILL$ translations of the sequents, when
provable. 
These results can be found in the PDFs at
\begin{center}
\url{https://github.com/carlosolarte/Linear-Logic-Prover-in-Maude}
\end{center}
in the directory \texttt{output}.

The proofs are grouped by intuitionistic problem, so one can compare the shape
and size of the original intuitionistic proofs and their linear versions
resulted from each translation. A summary of the results is presented in
Table~\ref{fig:comparison}. 

\newcommand{\fail}{\red{\textbf{\textsf{x}}}}
\newcommand{\tout}{\clock}

\begin{table}
\resizebox{\textwidth}{!}{
\begin{tabular}{ l | c |  c  | c |  c |  c |  c  | c |  c |  c |  c  | c |  c |  c | }
\cline{2-13}
& \multicolumn{4}{c|}{Encoding 0/1} &
\multicolumn{4}{c|}{Call-by-Name} & 
\multicolumn{4}{c|}{Call-by-Value}
\\\cline{2-13}
 & KLE & LCL & SYJ & SYN
 & KLE & LCL & SYJ & SYN
 & KLE & LCL & SYJ & SYN \\\hline
\multicolumn{1}{|l|} {Num. of Problems}&88&2&248&20&88&2&248&20&88&2&248&20\\\hline
\multicolumn{1}{|l|}{Unsolved (timeouts)}&11&2&209&9&12&2&203&9&9&2&209&9\\\hline
\multicolumn{1}{|l|}{Solved (Theorems)}&76&0&39&7&76&0&45&6&76&0&39&7\\\hline
\multicolumn{1}{|l|}{Non-Theorems}&1&0&0&4&0&0&0&5&3&0&0&0\\\hline
\multicolumn{1}{|l|}{Min Time}&17&-&18&16&17&-&17&16&16&-&17&17\\\hline
\multicolumn{1}{|l|}{Avg Time}&798.78&-&34039.10&105.81&250.43&-&20319.98&96.27&494.49&-&33884.36&77.18\\\hline
\multicolumn{1}{|l|}{Max Time}&41548&-&276886&458&4186&-&174778&383&24427&-&265778&321\\\hline

\end{tabular}
}

\caption{Comparison of translations. Times are measured  in {miliseconds}. Test run on a QEMU Virtual CPU, 2GHz, 64 bits, 6GB of RAM running Ubuntu. Timeout set to 5 minutes. 
\label{fig:comparison}} 
\end{table}

\subsection{Petri nets problems}\label{sec:petri}

Petri nets (see e.g.~\cite{DBLP:books/sp/Reisig85a}) are a model for distributed systems that generalize automata and hence also transition systems. Local
states in the net are called \emph{places} (denoted graphically as a circle) and
the global state is a  multiset of local states called a \emph{marking}. A
certain number $n$ of copies of a local state $s$ in a marking $M$ is  denoted
as $n$ tokens (graphically bullets) inside the circle $s$.  Transitions produce
and consume tokens, representing transformations in the  local states.

Abstractly, we can see a Petri net as a multiset rewrite system and thus, there
is a natural connection with the multiplicative fragment of linear
logic~\cite{petri-LL}. More precisely, a transition system   over the set of
places $S$ is a set of rewrite rules of the form $r: \bullet t \rhd  t\bullet$
where $\bullet t$ and $t\bullet$ are multisets on $S$, usually called preset and
postset. As expected, the rule $r$  is enabled in a marking (global state) $M$
if $\bullet t$ is a multisubset of $M$. Then, $M$ reduces to $M'$, notation $M
\rhd M'$, where $M' = M \setminus \bullet t \uplus t \bullet$. Let $\rhd^*$ be
the reflexive and transitive closure of $\rhd$. We say that $M'$ is reachable
from $M$ if $M \rhd^* M'$. 

Let $s\in S$ be a place and  $M,M'$ be a multisets on $S$. Then, we can encode
places, markings and rules as follows~\cite{petri-LL}:
\[
\begin{array}{lll l lll  l lll}
(s)^{pt} & = & \texttt{s} &\quad & (\{\})^{pt} &=& \one\\
 (\{s\} \uplus M)^{pt} &=& (s)^{pt} \otimes (M)^{pt} & \quad &
(r : M \rhd M') & = & !( (M)^{pt} \limp (M')^{pt})
\end{array}
\]
where $\texttt{s}$ is an atomic proposition. Given a 
set of transitions $\mathcal{R}$, the reachability problem whether $M \rhd^* M'$
can be encoded as the $\ILL$ sequent $\bigotimes\limits_{r\in \mathcal{R}}
(r)^{pt} \vdash (M)^{pt} \limp (M')^{pt}$.

The Petri Nets Repository\footnote{\url{https://pnrepository.lip6.fr/}} contains
a large collection of Petri nets in PNML (Petri Net Markup
Language\footnote{\url{http://pnml.lip6.fr/}}) syntax, an XML with tags for
places and transitions (or ``arcs''), and a way to represent the initial
marking. We wrote scripts that translate PNML into a simple Maude program that
can perform transitions in the net, transforming the markings. This program was
used to generate final markings of the Petri nets after 1, 5, 10, 20, 50 and 100
transitions. Each final marking, together with the Petri net specification, is a
reachability problem in linear logic, which was written in the ILLTP format
(described at the beginning of Section~\ref{sec:kleene}) using the translation
above. It is worth noticing that some nets are cyclic (deadlock free transition
systems), hence it is possible that the marking after $n$ transition can be also
obtained after $m<n$ transitions. 
This procedure allowed us to generate 3.137 new problems.

\section{Conclusion}\label{sec:conc}

In this work we proposed an initial set of problems as a library for
benchmarking linear logic based provers. Starting with the
$(\to,\land)$-fragment of Kleene's theorems, we generated 244 different $\ILL$
sequents using 4 automatic translations: {\em multiplicative}, {\em call-by-name}, {\em call-by-value} and {\em 0/1}. The first translation is the only one that does
not preserve provability. For each of those 22 $\ILL$ sequents that are not
provable via the multiplicative translation, we proposed an alternative provable
sequent with a small set of additives and bangs added. This makes these
particular sequents amenable to the use of all the power of focusing theorem
provers. In fact, the excess of bangs in formulas tends to neutralize the
efficacy of focusing, due to the positive/negative behavior of the exponentials. 
Thus our initial proposal for a suitable benchmark for $\ILL$ has 271 formulas,
testing aspects like provability and focusing.  

It is worth noticing that (1) we  include $\bot$ in the grammar of $\ILL$; (2)
all the sequents of our collection can also serve as tests in $\CLL$.  The
decision in (1) was motivated by the fact that the resulting sequents fall into
the multiplicative fragment of $\ILL$. However,  observe that one could clearly
exchange $\bot$ for $\zero$ in the multiplicative translation (that would not be
multiplicative anymore) and still obtain a significant set of 23 formulas not
provable via this new translation.  

We then moved in the direction of increasing the amount of formulas in our
library, still using the translation approach. For that, we took all the 274
formulas from the ILTP Library and applied the three sound translations from
$\IL$ to $\LL$ considered in this work. At this point, our library had already
1.093 formulas to be tested.  For an initial experiment with this first
proposed set of sequents, we implemented a focused-based prover for $\ILL$.
This allowed us to compare well known $\LL$ translations for $\IL$
formulas, not only considering the time needed for proving a sequent: our results give a comprehensive collection of proofs where it is possible to visualize the different shapes (and sizes) of such proofs. 

Finally, following Frank Pfenning's suggestion, we have generated automatically
several reachability problems from the large collection of Petri nets. This not
only moves us out of intuitionistic logic examples, but it also considers a
serious amount of problems that are actually used for testing Petri nets model
checkers. 

It should be noted that these are only the first steps on the direction of
benchmarking linear logic.  For future work, we intend to test different provers already available online and to propose a large collection of formulas dedicated to classical linear logic $\CLL$.

\bibliographystyle{eptcs}
\bibliography{references}

\appendix

\section{Multiplicative translation of Kleene's list}\label{app:mult}

\begin{minipage}{0.95\textwidth}
{\footnotesize
\begin{multicols}{2}
\begin{enumerate}\setlength\itemsep{0em}
\item $\vdash A\limp A$ (identity)
\item $A\limp B, B\limp C\vdash A\limp C$ (transitivity of implication)
\item $A\limp (B\limp C)\vdash B\limp (A\limp C)$ (exchange of premises)
\item $A\limp (B\limp C) \vdash A\otimes B\limp C$ (uncurrying)
\item $A\otimes B\limp C\vdash A\limp (B\limp C)$ (currying)
\item $A\limp B\vdash (B\limp C) \limp (A\limp C)$ (precomposing maps) 
\item $A\limp B\vdash (C\limp A) \limp (C\limp B)$ (post-composing maps) 
\item $A\limp B\vdash A\otimes C\limp B\otimes C$ (tensor is a bifunctor)
\item $A\limp B\vdash C\otimes  A \limp C\otimes B$ (tensor is a bifunctor)
\item $ A^{\bot}\not\vdash A\limp B$ 
\item $A\not\vdash A^{\bot}\limp B$ 
\item $B \not\vdash A\limp B$ 
but  (projection is non-linear)
\item 
$A\limp B\vdash  B^{\bot}\limp A^{\bot}$ (linear negation is contravariant)
\item  $A\limp  B^\perp\vdash  B^\perp\limp A^\perp$ \item  $A\limp B, B\limp A\vdash A\multimapboth B$  
\item  $A\multimapboth B\not \vdash A\limp B$ (cannot throw away $B\limp A$)
\item $A\multimapboth B\not \vdash B\limp A$ (cannot throw away $A\limp B$)
\item $A\multimapboth B, A\not \vdash  B$ 
\item $A\multimapboth B, B\not \vdash  A$ 
\item $\vdash A\multimapboth A$ 
\item $A\multimapboth B\vdash B\multimapboth A$ 
\item $A\multimapboth B, B\multimapboth C\vdash A\multimapboth C$  
 \item $A\limp (B\limp C),  A^{{\bot}{\bot}}, B^{{\bot}{\bot}} \vdash  C^{{\bot}{\bot}}$ 
\item $(A\limp B)^{{\bot}{\bot}}\vdash A^{{\bot}{\bot}}\limp B^{{\bot}{\bot}} $ (double negation is a functor)  
\item $ (A\limp B)^{{\bot}{\bot}}, (B\limp C)^{{\bot}{\bot}}\vdash  (A\limp C)^{{\bot}{\bot}}$ 
\item $\not\vdash  (A\otimes B)^{{\bot}{\bot}} \multimapboth  A^{{\bot}{\bot}}\otimes  B^{{\bot}{\bot}}$ 
\item $\not\vdash  (A\multimapboth B)^{{\bot}{\bot}} \multimapboth  (A\limp B)^{{\bot}{\bot}} \otimes  (B\limp A)^{{\bot}{\bot}}$ 
\item $A \multimapboth B \vdash (A \limp C) \multimapboth (B \limp C)$ 
\item $A \multimapboth B \vdash (C \limp A) \multimapboth (C \limp B)$ 
\item $A \multimapboth B \vdash (A \otimes C) \multimapboth (B \otimes C)$ 
\item $A \multimapboth B \vdash (C \otimes A) \multimapboth (C \otimes B)$
   \item $A \multimapboth B \vdash  A^{\bot} \multimapboth  B^{\bot}$  
\item $\vdash ((A \otimes B) \otimes C) \multimapboth (A \otimes (B \otimes C))$. 
\item $\vdash A \otimes B \multimapboth B \otimes A$
\item $\not\vdash A \otimes A \multimapboth A$ ($\otimes$ is not idempotent) 
\item $A \not\vdash (A \limp B) \multimapboth  B $ 
\item $B \not\vdash (A \limp B) \multimapboth  B$ 
\item $A^\perp \not\vdash (A \limp B)  \multimapboth A^\perp$ 
\item $ B^\perp \not\vdash (A \limp B)  \multimapboth A^\perp$ 
\item $B \not\vdash (A \tensor  B) \multimapboth A$ 
\item $B^\perp\not\vdash ((\bang A\tensor B)\limp B)\tensor (B\limp (A\tensor B))$ 
\item $\vdash A \limp A^{\perp\perp}$ 
\item $\vdash A^{\perp\perp\perp}  \multimapboth A^\perp$ 
\item  $\vdash (A \otimes A^\perp)^\perp$
\item $\not\vdash (A\multimapboth A^\perp)^\perp$ 
\item $\not\vdash (((A\limp\zero)\limp\zero) \limp A)^{\perp\perp}$ 
\item $\not\vdash A \otimes (B \otimes B^\perp) \multimapboth (B \otimes B^\perp)$ 
\item $\vdash (A \limp B) \limp (A \otimes B^\perp)^\perp$ 
\item $\vdash (A \limp B^\perp) \multimapboth  (A \otimes B)^\perp$
\item $\vdash  (A \otimes B)^\perp \multimapboth (A^{\perp\perp} \limp B^\perp)$ 
\item $ B^{\perp\perp} \limp B \vdash (A^{\perp\perp} \limp B) \multimapboth (A \limp B)$ 
\item $ B^{\perp\perp} \limp B \vdash  (A \limp B) \multimapboth  (A \otimes B^\perp)^\perp$
\item $\vdash ( A^{\perp\perp} \limp B) \limp (A \otimes  B^\perp)^\perp$ 
\item $\vdash A \otimes B \limp  (A \limp B^\perp)^\perp$ 
\item $\vdash A \otimes B^\perp \limp (A \limp B)^\perp$ 
\item $\vdash (A^{\perp\perp} \otimes B) \limp  (A \limp B^\perp)^\perp$  
\item $\not\vdash (A^{\perp\perp} \otimes B^\perp) \multimapboth  (A \limp B)^\perp$ 
\item $\not\vdash (A \limp B)^\perp\multimapboth   (A \otimes B^\perp)^{\perp\perp}$ 
\item $\not\vdash (A \limp B)^{\perp\perp} \multimapboth  (A \otimes B^\perp)^\perp$
\item $\vdash  (A \otimes B^\perp)^\perp \multimapboth (A \limp  B^{\perp\perp}) $ 
\item $\vdash (A \limp  B^{\perp\perp}) \multimapboth (A^{\perp\perp} \limp B^{\perp\perp})$
\end{enumerate}
\end{multicols}
}
\end{minipage}

\end{document}